\documentclass[12pt,preprint]{aastex}
\usepackage{emulateapj5}

\newcommand{\gsim}{\raisebox{-0.3ex}{\mbox{$\stackrel{>}{_\sim} \,$}}}
\def\gta{\ifmmode {\mathbin{\lower 3pt\hbox   
    {$\,\rlap{\raise 5pt\hbox{$\char'076$}}\mathchar"7218\,$}}}
    \else {${\mathbin{\lower 3pt\hbox
    {$\rlap{\raise 5pt\hbox{$\char'076$}}\mathchar"7218\,$}}}
    $}\fi}
\def\lta{\ifmmode {\,\mathbin{\lower 3pt\hbox   
    {$\,\rlap{\raise 5pt\hbox{$\char'074$}}\mathchar"7218\,$}}}
    \else {${\mathbin{\lower 3pt\hbox
    {$\rlap{\raise 5pt\hbox{$\char'074$}}\mathchar"7218\,$}}}
    $}\fi}

\shorttitle {Discovery of a neutron star in A1744--361} 
\shortauthors {Bhattacharyya et al.}

\begin{document}

\title {The Discovery of a Neutron Star with a Spin Frequency of 530 Hz in A1744--361}

\author {Sudip Bhattacharyya\altaffilmark{1,2}, Tod
E. Strohmayer\altaffilmark{2}, Craig B. Markwardt\altaffilmark{1,2}, and
Jean H. Swank\altaffilmark{2}}

\altaffiltext{1}{Department of Astronomy, University of Maryland at
College Park, College Park, MD 20742-2421}

\altaffiltext{2}{X-ray Astrophysics Lab,
Exploration of the Universe Division,
NASA's Goddard Space Flight Center,
Greenbelt, MD 20771; sudip@milkyway.gsfc.nasa.gov, 
stroh@clarence.gsfc.nasa.govi, craigm@milkyway.gsfc.nasa.gov,
swank@milkyway.gsfc.nasa.gov}



\begin{abstract}

We report the detection with the Rossi X-ray Timing Explorer (RXTE)
Proportional Counter Array (PCA) of 530 Hz burst oscillations in a
thermonuclear (type I) burst from the transient X-ray source
A1744--361. This is only the second burst ever observed from this
source, and the first to be seen in any detail. Our results confirm
that A1744--361 is a low mass X-ray binary (LMXB) system harboring a
rapidly rotating neutron star.  The oscillations are first detected
along the rising edge of the burst, and they show evidence for frequency
evolution of a magnitude similar to that seen in other burst
sources. The modulation amplitude and its increase with photon energy
are also typical of burst oscillations. The lack of any strong
indication of photospheric radius expansion during the burst suggests
a 9 kpc upper limit of the source distance. We also find energy
dependent dips, establishing A1744--361 as a high inclination, dipping
LMXB. The timescale between the two episodes of observed dips suggests
an orbital period of $\sim 97$ minutes. We have also detected a 2$-$4 Hz
quasi-periodic-oscillation (QPO) for the first time from this
source. This QPO appears consistent with $\sim 1$ Hz QPOs seen from
other high-inclination systems. We searched for kilohertz QPOs, and
found a suggestive $2.3\sigma$ feature at 800 Hz in one
observation. The frequency, strength, and quality factor are consistent
with that of a lower frequency kilohertz QPO, but the relatively low
significance argues for caution, so we consider this a tentative
detection requiring confirmation.

\end{abstract}


\section {Introduction} \label{sec: 1}

The X-ray transient A1744--361 was discovered by Ariel V in 1976, when
the source was in outburst (Davison et al. 1976; Carpenter et
al. 1977). No optical counterpart was identified (Burnell \&
Chiappetti 1984), and no thermonuclear (type I) X-ray bursts were
reported.  The source was observed to be in outburst again 13.5 years
later in August 1989 (in't Zand 1992; in 't Zand 2004). This time it
was detected by the Coded Mask Imaging Spectrometer (COMIS/TTM; in't Zand 1992)
onboard the `Mir' space station, and the source was at a relatively constant
level of $\sim 60$~mCrab (about three times fainter than in 1976; see
in't Zand 1992). It was at least 10 times weaker (and undetectable)
five months earlier and one month later (in 't Zand 2004). The source
spectrum was indicative of a soft transient (in't Zand 1992).
Emelyanov et al. (2001) analyzed data from COMIS/TTM and found 33
likely type I bursts from several X-ray sources in the field
containing A1744--361. They found a burst from the Aug 23, 1989 data
from A1744--361. The burst was detected with 8 s time resolution, by
comparison of the source flux during the likely burst and the
session-averaged source flux. No burst profile or detailed information
about this burst was given. Moreover, Emelyanov et al. (2001) made a
classification of the observed bursts as type I (thermonuclear) based
only on their identification with known classical bursters, while at
the time A1744--361 was not known to be a burster. However, they
considered this burst to be the first detection of a type I X-ray burst
from this source, and suggested that A1744--361 contains an accreting
neutron star.

Since 2003, A1744--361 has been in outburst every year, detected by
RXTE (ASM and PCA), Chandra, and INTEGRAL (Remillard et al. 2003;
McClintock et al. 2003; Torres et al. 2004; Markwardt \& Swank 2004;
Grebenev et al. 2004; Swank \& Markwardt 2005). The detections of radio
and optical counterparts of this source have also been reported (Rupen
et al. 2003; Steeghs et al. 2004). However, the X-ray detections (in 2003
\& 2004) did not show any type I X-ray bursts.

Discovery of type I bursts from an X-ray source can conclusively
classify it, as these bursts are produced by thermonuclear burning of
matter accumulated on the surfaces of accreting neutron stars
(Joss 1977; Lamb, \& Lamb 1978). Moreover, thermonuclear bursts
are observed only from LMXBs (Strohmayer \& Bildsten 2003; Liu, van
Paradijs, \& van den Heuvel 2001). By analyzing 2005 RXTE PCA data,
we have found a type I burst from this source, which confirms the
earlier conclusion of Emelyanov et al (2001).

We have also discovered millisecond period brightness oscillations
with a frequency of $\approx 530$ Hz during this burst. These
oscillations are produced by an asymmetric brightness pattern on the
stellar surface that is modulated by rotation of the star (Chakrabarty
et al. 2003; Strohmayer \& Bildsten 2003).  Therefore, the burst
oscillation frequency is identical to, or very close to the stellar
spin frequency. Hence, our detection of $\sim 530$~Hz burst
oscillations establishes the spin frequency of the neutron star in
A1744--361.

Analysing the 2003 RXTE PCA data, we have discovered two episodes of
energy dependent dips in the X-ray flux from A1744. The time interval
between the two sets of dips in two successive RXTE orbits indicates
that the orbital period of A1744--361 is $97\pm22$~min. We also report
the discovery of a low frequency quasi-periodic-oscillation (QPO;
$\sim 3$~Hz), as well as an indication of a kHz QPO $(\sim
800$~Hz) from this source.  In \S~2, we describe the analysis of the
data, and present our results. In \S~3, we discuss the implications of
our findings.

\section {Data Analysis and Results} \label{sec: 2}

We analyzed RXTE PCA data from the transient source A1744--361 when it
was in outburst in 2003, 2004, and 2005. 
The duration (RXTE proposal number, RXTE observation duration)
of these outbursts were $\sim 1$ month (P80431, $\sim 39.1$~ks), 
$\sim 20$ days (P90058, $\sim 1.9$~ks), and $\sim 40$ days (P91050, $\sim 14$~ks)
respectively. We analyzed all the $\sim 55$~ks of data, and
found a single thermonuclear X-ray burst (July 16, 2005). The rise
time of the burst is $\sim 1$~s, while the decay time is $\gsim 10$~s
(see Fig. 1). The peak count rate of the burst was $\gsim 6000$ with 3
PCUs operating. We created burst profiles for different energy ranges,
and also hardness profiles for several pairs of energy ranges. We also
performed a time resolved spectral analysis by fitting blackbody
spectra (for the fixed hydrogen column density $0.79\times10^{22}$~cm$^{-2}$; 
in't Zand 1992) 
through the burst. The lack of any significant drop in the
blackbody temperature correlated with an increase in the blackbody
normalization (ie. radius), leads us to conclude that this is not a
photospheric radius expansion (PRE) burst. 
The burst had a peak flux in the 2 - 20 keV band of $1.9 \times 10^{-8}$ ergs cm$^{-2}$
s$^{-1}$. Considering this flux to be less than the Eddington flux 
(Shapiro \& Teukolsky 1983),
the upper limit to the source distance is $\approx 9$~kpc, for a 1.4
$M_\odot$ neutron star mass, ionized hydrogenic accreted matter, and
isotropic emission from the star. However, this upper limit will increase 
for lesser hydrogenic abundance, and will
decrease if the gravitational redshift due the neutron star is
considered (Galloway et al 2005).

In order to search for oscillations during the burst, we calculated
power spectra with a Nyquist frequency of 2048 Hz for 4 s intervals
starting from the burst onset, using 125 $\mu$s event mode data. We
found a candidate peak at $\sim 530$~Hz in the first such
spectrum. The main panel of Fig. 1 shows the 2 - 60 keV burst profile
and the 4 s interval used to compute the power spectrum. The peak was
resolved, so lowering the frequency resolution by averaging adjacent
fourier bins improved the signal to noise ratio.  At 2 Hz resolution
we found a peak power of 10.3. The inset panel of Fig. 1 shows the 2
Hz resolution power spectrum. The probability of obtaining a power
this high in a single trial from the expected $\chi^2$ noise
distribution (16 dof) is $\approx 6.13 \times 10^{-11}$. Multiplying
by a conservative trials penalty of 8192, the number of frequency bins
in the original spectrum, we arrive at a significance of $5.02 \times
10^{-7}$, which indicates a strong detection.

To get a rough idea about possible frequency evolution, we next
calculated a dynamic $Z^2$ power spectrum (Strohmayer \& Markwardt
1999).  We used 1 s intervals to compute $Z^2$ power spectra, and
started a new interval every 1/8 s. The corresponding power contours
(see Fig. 2) are associated primarily with the rising portion of the
burst profile, and the time evolution suggests that the oscillation
frequency increases somewhat during the burst rise to peak.  These
properties are fairly typical of the behavior of oscillations seen in
other burst sources, giving us even added confidence in the detection.
The highest amplitude in the $> 3$ keV band in a 1 s interval during
the oscillation is 10.3 \% (rms). The amplitude increases with energy,
reaching 15\% for photons above 8 keV.  This behavior is also fairly
typical of burst oscillations (Strohmayer et al. 1997; Muno, Ozel \&
Chakrabarty 2003).  The pulse profile is sinusoidal, with no
indications of significant harmonic structure.

We searched all the data for QPOs, and found a $\sim 3$~Hz QPO in the
April, 2004 data. We divided 750 s of data from the ObsID
90058-04-01-00 into $M$ (= 75) equal segments of 10 s duration, and
we calculated fast Fourier transforms on each time segment. In order to
reduce the noise, the resulting power spectra were averaged, and $W$
(= 8) consecutive frequency bins were combined (making the frequency
resolution 0.8 Hz). The upper figure of panel {\it a} of Fig. 3 shows this power spectrum,
which clearly depicts a QPO of quality factor $(Q) \sim 2$, at $\sim
3.5$~Hz (rms amplitude $\sim 5$\%).  For computing the significance of this QPO, we fitted the
spectrum with a model, and minimized the corresponding $\chi^2$ to get
the best fit parameter values. Then we divided the power spectrum by
the best fit model, and multiplied by 2 (see the upper figure of panel {\it b}, Fig. 3),
in order to have the noise distributed as $\chi^2$ with $2MW$ degrees
of freedom. The peak power (2.45) of the QPO, therefore, has the
single trial significance of $1.25\times10^{-7}$. As low frequency
QPOs are searched up to 100 Hz, multiplying by the number (= 125) of
trials, we get a significance of $1.56\times10^{-5}$, which implies a
$\sim 4.3\sigma$ detection. We also note that this QPO seems to shift
by $\sim 1$~Hz in $\sim 40$ hours, as there is an indication of a
$\sim 2.5$~Hz QPO (with significance $\sim 2.6\sigma$, and rms amplitude $\sim 3$\%; 
lower figures of panels {\it a} \& {\it b} of Fig. 3) in the ObsID
90058-04-02-00 (Apr 10, 2004).

We searched for kHz QPOs in the whole data set, and found a tentative
indication of a 800 Hz QPO in the ObsID 90058-04-02-00. The
significance of this possible QPO is only $\sim 2.3\sigma$. However,
its centroid frequency $(\sim 800$~Hz) and high $Q$-value $(\sim
62.5)$ are consistent with those of lower kHz QPOs observed from other
sources. The inferred amplitude (rms) of $\sim 6$\% is also consistent with
lower frequency kHz QPOs in other sources. 

We discovered intensity dips in the ObsID 80431-01-02-00 (Nov 17,
2003).  The observed durations of the dips are between 5 s and 25 s.
These dips appear only in the softer energy bands (see Fig. 4), which
indicates that they are caused by partial obstruction of the central
X-ray source by structures (above the equatorial plane) created within
the accretion flow (White \& Swank 1982; Jonker et al. 2000). We found
two sets of dips in two subsequent data segments.  A particular set of
dips may be caused by cold clouds (or other structures) distributed in
a range of azimuthal angles above the accretion disk plane (see for
example, Frank, King, \& Lasota 1987). The time separation $(97$~min)
between the two subsequent sets of dips gives an estimate of the
orbital period of the binary system (White \& Swank 1982). As the
width of each set of dips introduces an uncertainty, we suggest that
the orbital period of A1744--361 is $97\pm22$~min. However, we note
that the actual orbital period may be half (as we may have missed a
set of dips because of the gap between two data segments), or twice
this value (as secondary dips may occur in between two primary sets of dips;
Smale et al. 1989).

\section {Discussion and Conclusions}

The X-ray transient A1744--361 has been an elusive source, observed
only twice before 2003 (in 1976 and in 1989) with gaps of many
years. As a result, detailed information about this source was
lacking, except the strong indication (from an putative type I X-ray
burst; Emelyanov et al. 2001) that A1744--361 is a neutron star LMXB.
Since 2003, the source has shown outbursts every year. Analyzing 2005
RXTE PCA data, we found a thermonuclear burst from this source with
certainty (as no other known candidate source was in the PCA field of
view). We also discovered millisecond period brightness
oscillations during this burst. From these results we confirm that
this source is an LMXB harboring a neutron star. We infer the spin
frequency of the neutron star to be $\sim 530$~Hz from the burst
oscillations. This value is consistent with the observed spin
frequencies of the neutron stars in LMXBs, that range from 45 Hz to
619 Hz (Strohmayer \& Bildsten 2003; Kaaret et al. 2005).
More observations of bursts with oscillations from this source will be 
important, because the coherent addition of oscillation signals from
several bursts may lead to a significant detection of harmonic power
(Bhattacharyya \& Strohmayer 2005b), which will be very useful for
constraining neutron star mass and radius, and hence for understanding
the dense cold matter at the stellar core (Bhattacharyya et al. 2005).
The oscillation frequency of the observed burst seems to increase
somewhat with time, and such evolution during burst rise can be
used to understand the spreading of the thermonuclear flames on the
neutron star surface (Bhattacharyya \& Strohmayer 2005b; 2005c). Such
an understanding (Spitkovsky, Levin, \& Ushomirsky 2002; Bhattacharyya
\& Strohmayer 2005a; 2005c; 2005d) may be useful for constraining stellar
surface and structure parameters.

We also discovered intensity dips in the soft X-ray band from
A1744--361, but have not observed any eclipses. This establishes that
A1744--361 is a `pure' dipper, which suggests that the observer's
inclination angle $i$ is in the range $\sim 60^{\rm o} - 75^{\rm o}$
(Frank et al. 1987).  Two subsequent sets of dips also suggest that
the orbital period of this binary system is $P = 97\pm22$~min, which
is consistent with the observed orbital periods of other compact LMXB
systems harboring rapidly rotating neutron stars.  Using this orbital
period and the equation $P \approx 9^{\rm h} (R_{\rm
comp}/R_\odot)^{3/2} (M_\odot/M_{\rm comp})^{1/2}$ (assuming that
the donor star fills its Roche lobe; Bhattacharya \&
van den Heuvel 1991), we can draw a curve in the radius-mass plane of
the secondary companion star. Here, $9^{\rm h}$ is 9 hours, $R_{\rm
comp}$ \& $M_{\rm comp}$ are the companion radius and mass, and
$R_\odot$ \& $M_\odot$ are the solar radius and mass. This radius-mass
relation, and the obtained range of $i$, would allow us to determine
the nature of the companion star, if either the pulsar mass function
or the projection of the orbital velocity of the neutron star along
the line of sight were known.  However, even with the available
information, we can make some useful comments.  To do this, we keep in
mind that for a given nature and mass of the companion, if $R_2$ is
the normal stable radius, then the actual radius $R_{\rm comp}$ may be
greater than or equal to $R_2$ (due to bloating by X-ray heating), but
can not be less than it. Consequently, if $M_{\rm comp} > 0.18
M_\odot$, the companion can not be a hydrogen main sequence star. As
for $M_{\rm comp} > 0.18 M_\odot$ a degenerate or helium main
sequence companion would have to be bloated by at least several or
many times its original volume, it is likely that $M_{\rm comp} < 0.18
M_\odot$. This suggests that the companion star in A1744--361 is
likely to be a slightly bloated brown dwarf or hydrogen main sequence
star (see Bildsten \& Chakrabarty 2001). However, we note that if 
the actual orbital period is double of the proposed one, the companion
star may be a hydrogen main sequence star (with the upper limit of 
$M_{\rm comp} \approx 0.35 M_\odot$), or a brown dwarf, while an 
orbital period half the proposed value will allow for a brown dwarf and
a hydrogen main sequence star (for $M_{\rm comp} < 0.1 M_\odot)$, as well as
a bloated helium main sequence star with larger mass (may be $> 0.6 M_\odot)$.

We have also discovered low frequency QPOs $(\sim 3$~Hz) for the first
time from A1744--361. This is consistent with the fact that this
source is a dipper, as such QPOs have been observed from other dipping
LMXBs (Jonker et al. 2000). It has been proposed that these QPOs are
caused by the partial obscuration of the central X-ray source by a
nearly opaque or gray medium in or on the accretion disk (Jonker et
al. 2000) that requires a high system inclination angle $(i)$, which
is the case for a dipper. We also report an indication of a kHz QPO
from this source, which if confirmed, would likely be a lower kHz
QPO. If confirmed, this will be the first kHz QPO observed from
A1744--361.

\acknowledgments

{}

\clearpage
\begin{figure}
\epsscale{1.0}
\plotone{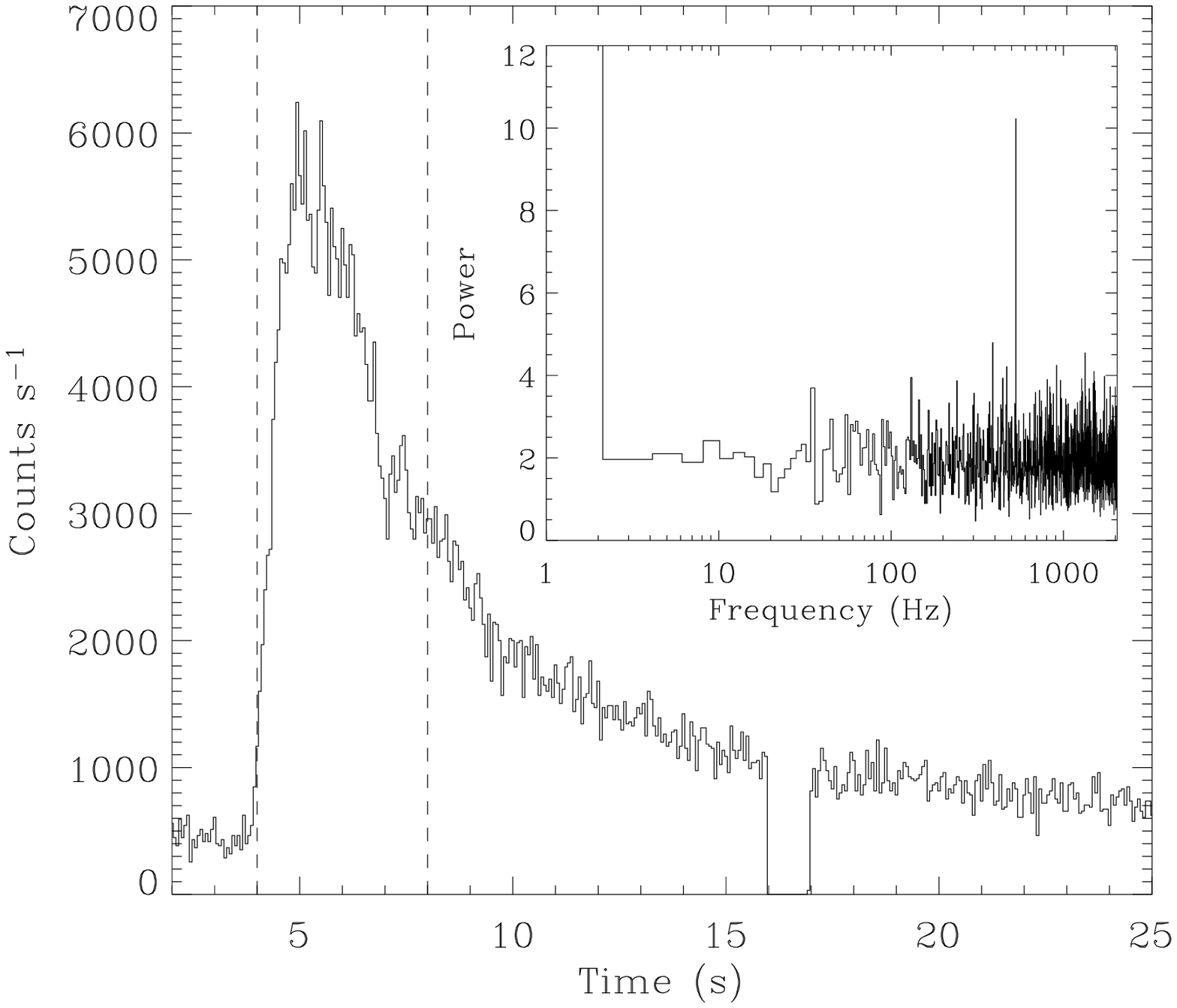}
\caption {Thermonuclear X-ray burst from A1744--361. The main panel
shows the PCA countrate (three detectors operating) profile in the 2 - 60
keV band. The inset panel shows the power spectrum of the first 4 s
interval (marked with vertical dashed lines in the main panel),
rebinned to 2 Hz resolution. The peak near 530 Hz stands out well
above the noise level.}
\end{figure}

\clearpage
\begin{figure}
\epsscale{1.0}
\plotone{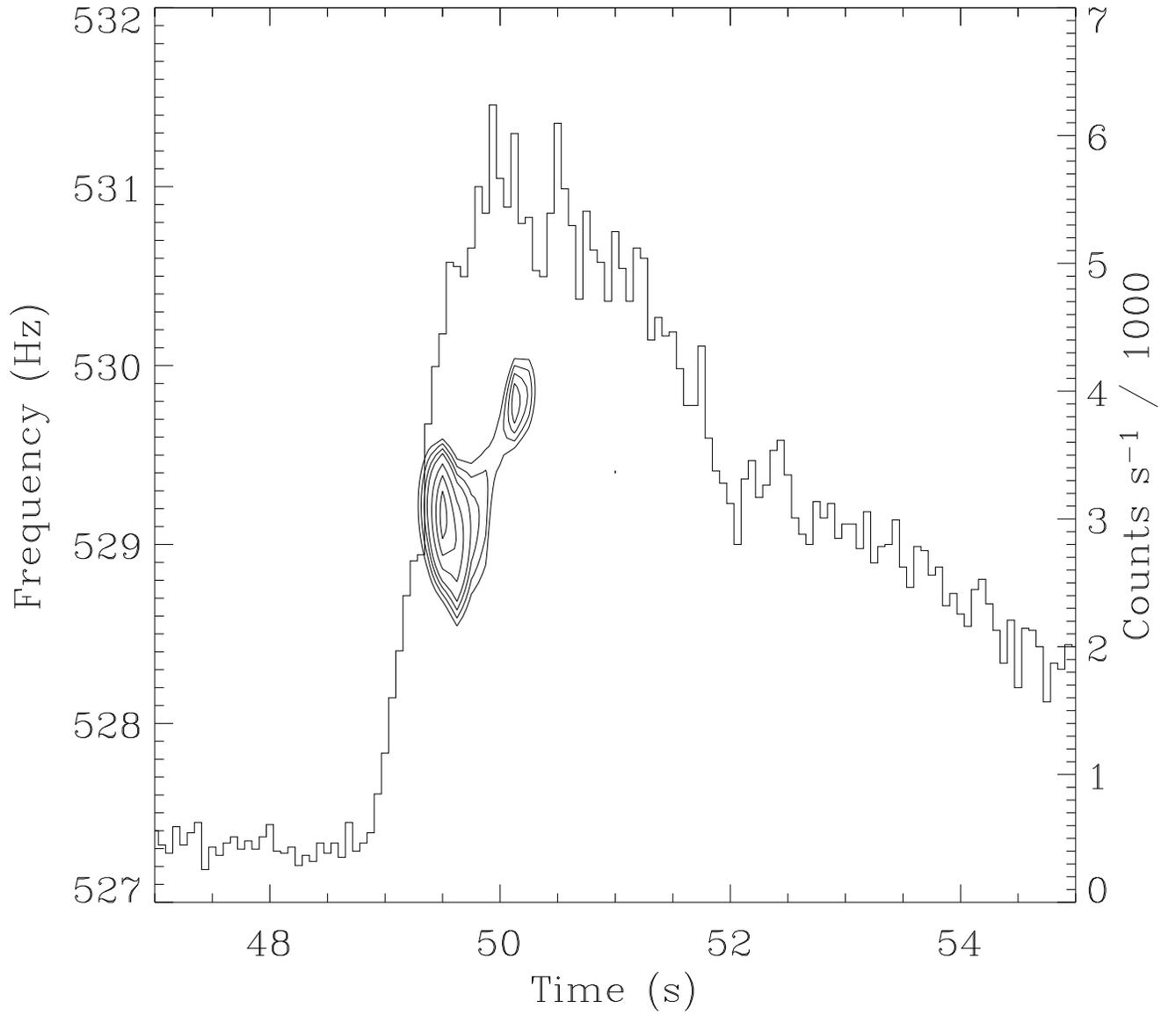}
\caption {Dynamic power spectrum of the burst from A1744--361. Power
spectra were computed using the $Z^2$ statistic in 1 s intervals,
with a new interval starting every 1/8 s.  The energy band was 2 - 60
keV. Contour levels of $Z^2 = 20, 22, 24, 26, 30, 34,$ and 38 are
shown.  The peak $Z^2$ power was 41.1.  }
\end{figure}

\clearpage
\begin{figure}
\epsscale{0.8}
\plotone{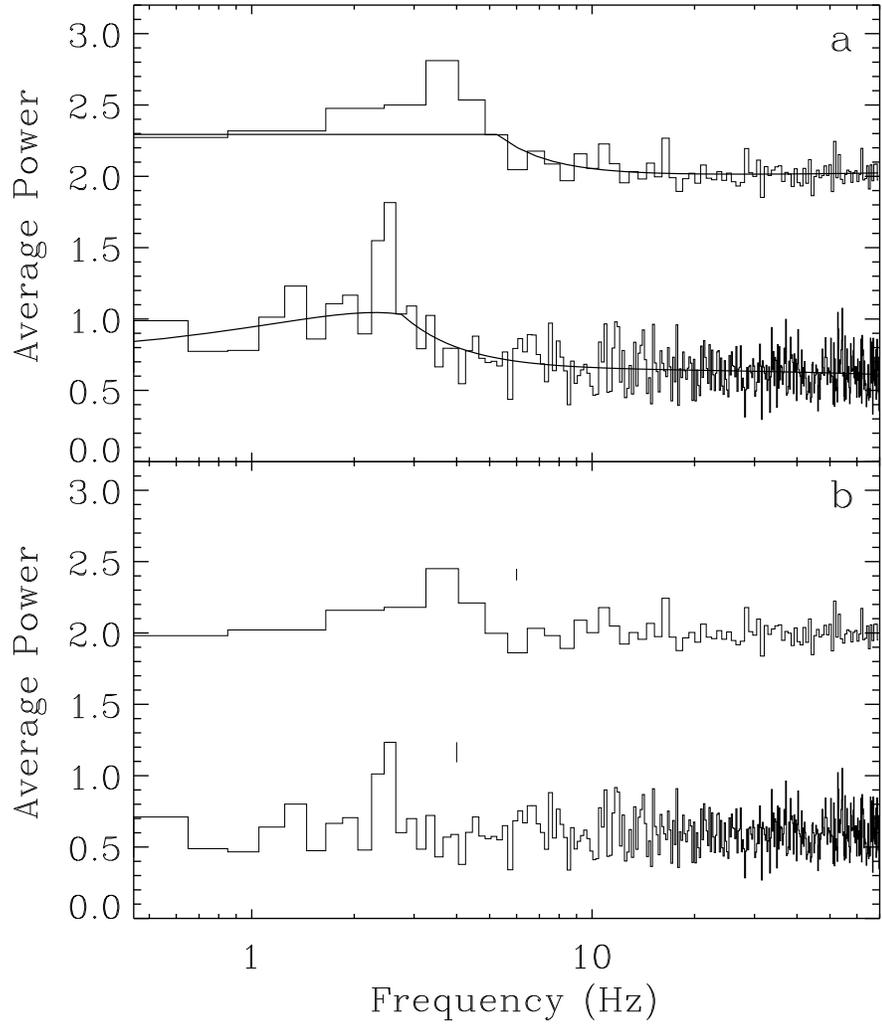}
\caption {Low frequency quasi-periodic-oscillations (QPOs) from
A1744--361. Panel {\it a} shows the power spectra for the ObsID
90058-04-01-00 (upper figure), and the ObsID 90058-04-02-00
(lower figure; power shifted by $-1.4)$. The solid curves give the best fit models of the
continua. In panel {\it b}, each of these power spectra has been divided by the
best fit model, and then multiplied by 2. The isolated vertical lines
give the corresponding size of the $1\sigma$ error. For these panels, powers are
calculated for 10 s data segments, and averaged over $M$ such
segments. 
For the ObsID 90058-04-01-00, $M = 75$, and frequency resolution is 0.8 Hz,
while these numbers are $100$ and 0.2 Hz for the ObsID 90058-04-02-00.}
\end{figure}

\clearpage
\begin{figure}
\epsscale{0.8}
\plotone{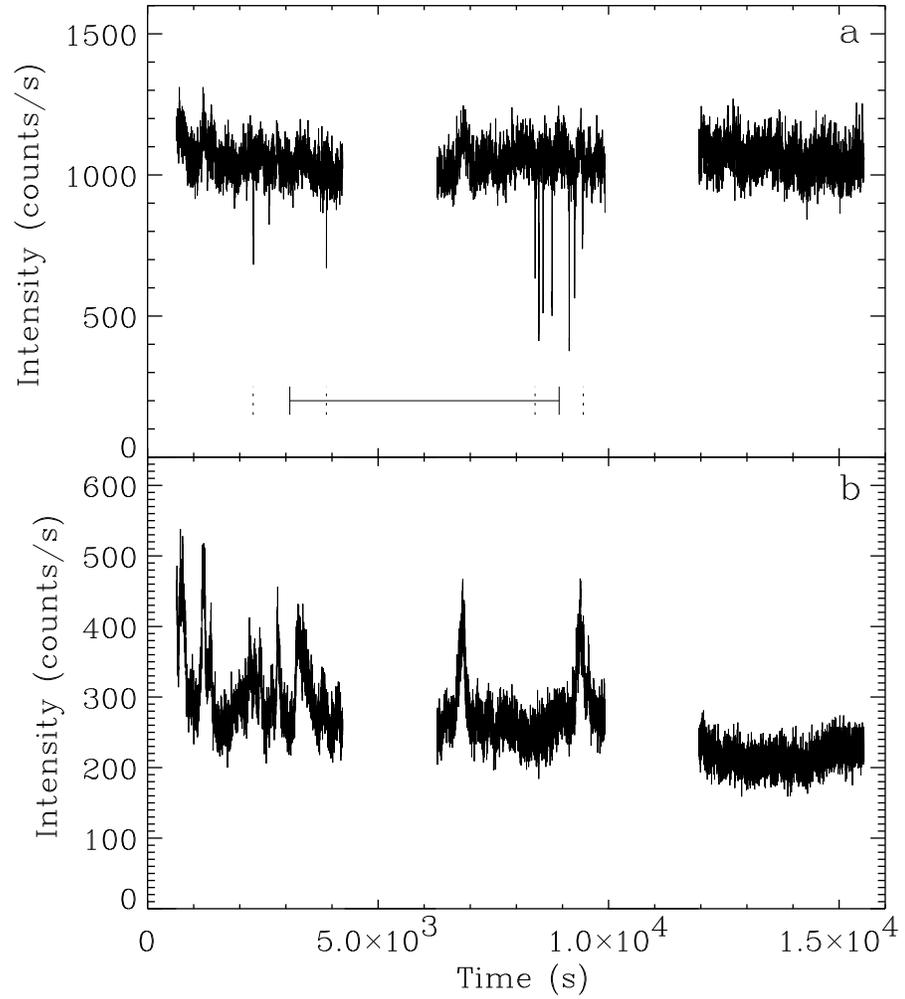}
\caption {Energy dependent dips in the lightcurves from
A1744--361. Both the panels are for three time segments of the ObsID
80431-01-02-00, the panel {\it a} is for the PCA channel range $0-13$,
while the panel {\it b} is for the channel range $14-63$.  Dips due to
obstruction of the central source by the structures in accretion flow
appear only in the soft energy lightcurves (panel {\it a}). The solid
horizontal line in the panel {\it a} gives the approximate orbital
period, and the dotted vertical lines give the corresponding
uncertainties on both sides. }
\end{figure}

\end{document}